\documentclass[10pt,twoside]{article}

\usepackage{natbib}
\def\apj{\textit{ApJ}} 
 
\def\aap{\textit{A\&A}} 

\def\mnras{\textit{MNRAS}}
\def\solphys{\textit{Solar Phys.}}
\def\araa{\textit{ARA\&A}}
\def\ssr{\textit{Space Science Reviews}}
\def\pasj{\textit{PASJ}}
\def\aapr{\textit{A\&A Reviews}}

\usepackage{graphicx}
\begin{document}

\thispagestyle{empty}

\vspace{-1cm} 
\noindent Submitted to the Hinode 4 proceedings:\\
``Hinode-4: unsolved problems and recent insights'', 11-15 October 2010 (Palermo), \\
editors: L.R.~Bellot Rubio, F.~Reale, and M.~Carlsson.

\begin{center}
{\large\bf On the Formation of Penumbrae as Observed with the German VTT, SOHO/MDI, and SDO/HMI}

{Rolf Schlichenmaier, Reza Rezaei, Nazaret Bello Gonz\'alez}

{Kiepenheuer-Institut f\"ur Sonnenphysik, Freiburg, Germany}

\end{center}

\paragraph{Abstract:}
Solar magnetic fields are generated in the solar interior and pop up at the solar surface to form active regions. As the magnetic field appears on the surface, it forms various structures like small magnetic elements, pores, and sunspots. The nature of these formation processes is largely unknown. In this contribution we elaborate on the formation of sunspots and particularly on the formation of penumbrae.

We report on observations that we obtained at the German VTT\footnote{VTT: The German \underline{V}acuum \underline{T}ower \underline{T}elescope is located in the `Observatorio del Teide' in Tenerife.} on July 4, 2009 on the formation of the spot in AR 11024. This data set is accomplished with data from SOHO/MDI\footnote{The \underline{M}ichelson  \underline{D}oppler  \underline{I}nterometer measures intensity images, velocity maps and maps of the longitudinal magnetic field on board the  \underline{SO}lar  \underline{H}eliospheric  \underline{O}bservatory.} which offers an entire time coverage. Moreover, the evolution of AR 11024 is compared with a particular event of penumbra formation in AR 11124 around November 13, 2010, using intensity images from SDO/HMI\footnote{The  \underline{H}elioseismic and  \underline{M}agnetic  \underline{I}mager onboard the  \underline{S}olar  \underline{D}ynamic  \underline{O}bservatory maps 4 Stokes parameters of Fe I 617.3\,nm with five spectral points of the entire solar disk.}.

We conclude that two processes contribute to the increase of the magnetic flux of a sunspot: (1) merging pores, and (2) emerging bipoles of which the spot polarity migrates towards and merges with the spot. As the penumbra forms the area, magnetic flux, and maximum field strength in the umbra stay constant or increase slightly, i.e. the formation of the penumbra is associated with flux emergence and an flux increase of the proto-spot. If two pores merge or if a pore merges with a proto-spot a light bridge is created. This initial light bridge dissolves in the further evolution. 


\section{Introduction}

The origin of magnetic fields in the Universe is a fundamental and unsolved problem. The modern Universe shows surprisingly strong and coherent magnetic fields on scales from a few kilometers to millions of parsecs \citep{rbeck2009}. This implies a set of ubiquitous processes that have amplified and organized the weak primordial field, and have maintained these fields over cosmic times. This process may have originally occurred in individual stars, with supernovae and outflows then diffusing the field out to large scales \citep{parker1979}. 

Magnetic fields hold the key to understand many long-standing problems in plasma physics and astrophysics. The large-scale coherent magnetic fields seen in galaxies and clusters provide tests of dynamo theory \citep{rbeck_etal_96}. 
However due to poor spatial resolution in galactic and extragalactic observations, the Sun stayed as a unique laboratory to study magnetic fields. 
An understanding of the structure and evolution of the fields is a necessity for solving the problem of magnetic field generation in the solar interior. 

The solar magnetic field is organized in two categories: large scale magnetic structures, known as active regions (AR), and small scale magnetic fields, mainly in the inter-network regions. An enigmatic property of the organized field on the Sun is that despite orders of magnitudes difference in length scale, total flux, and lifetime of the two categories, the difference in the field strength is less than a factor 10 \citep{solanki_etal_06}. 

It is commonly believed that the (large-scale) magnetic fields are generated in the solar interior, transferred through the convection zone, and finally emerge into the photosphere. 
The solar photosphere is convectively stable. That causes troubles for the large flux tubes to penetrate this layer~\citep{vandriel_culhane09,lites_09_ar}. Therefore, large flux tubes fragment into many smaller flux tubes. Finally, such smaller flux tubes reach the surface as emerging bipoles and are observed as elongated granules \citep{zwaan1985,strous+zwaan1999,otsuji_etal_07, schlichenmaier+al2010a}.  

Earlier studies of the emergence of magnetic flux in active regions lead to prominent findings, namely Hale's law \citep{hale_nicholson_1925}, the butterfly diagram \citep{carrington} and Joy's law \citep{hale_etal1919}. Despite these important findings, details of modern observations demand a more elaborate explanation, e.g, the properties of an $\Omega$-loop like asymmetry, tilt, and twist \citep{driel_etal90, kosovichev09}. 
The emerged flux tubes have a common root, causing them to merge 'again' in the solar photosphere~\citep{spruit1981}. As a result of coalescence of the emerged flux, pores and sunspots form in ARs~\citep{zwaan1992}. 

Motivated by the buoyant flux tube model of \cite{parker1955}, there have been attempts to simulate the flux emergence in the solar atmosphere and explain their observed properties \citep{fan_etal93, moreno+al1994, caligari+moreno+schuessler1995, caligari+al98, abbet_etal00}. Recent simulations reproduce the organization of flux at the emergence site, namely the elongated granules~\citep{cheung_etal07, cheung_etal08, tortosa_etal09}. There are also attempts to study implications of flux emergence in the chromosphere and the corona \citep{sykora_etal09, fang_etal10}. 

Sunspots are the largest magnetically driven phenomena in the solar photosphere \citep{solanki03r, weiss_06, schlichenmaier2009}. Although their average properties \citep[e.g.,][]{keppen_martinez_96} and decay mechanisms~\citep[e.g.,][]{petrovay_vandriel_97} are well studied, their formation remained obscur.
There are many observations of flux emergence. Nevertheless, penumbral formation is poorly understood, mainly due to the lack of observations and challenge of simulations. From the observational side, \cite{leka_sku98, yang_etal03}, and \cite{schlichenmaier+al2010a, schlichenmaier+al2010b} (paper I and II, respectively, hereafter) are the only reports about the moments of penumbra formation. \cite{leka_sku98} and \cite{yang_etal03} observed the formation of only a few penumbral filaments. In none of them, a real sunspot is formed. Paper I and II present the first spectro-polarimetric observation in which a pore converts into a fully developed spot with a penumbra.

In a simulation of sunspot formation, parameters like plasma-$\beta$ or mass density span many orders of magnitudes. Hence, to perform realistic simulations is a demanding task which was yet not addressed completely.
\cite{cheung_etal10}, in an interesting attempt, present a simulation in which coalescence of elongated granules form a spot without a penumbra. In this simulation, the rise of a big flux tube, initially located at a depth of 7.5\,Mm, forms elongated granules at the $\tau\,=\,1$ surface in agreement with observations. Since the field lines have a common root (for each polarity), they converge and assemble two spots. Although this simulation fails to form a penumbra, it reproduces a significant part of the flux emergence phenomena in active regions.

In this contribution we report on observational characteristics of structure formation in active regions. In Sect.~\ref{sec:11024} we summarize the present status of our data analysis and of our results concerning AR 11024. This includes spectro-polarimetric VTT data at high resolution as well MDI/SOHO data which offers a complete time coverage and a larger field of view.  In Sect. \ref{sec:11124} we take advantage of the high resolution of HMI to document and compare the evolution of AR 11124 with the evolution of AR 11024. In Sect.\ref{sec:summary} we summarize our findings and draw conclusions.

\begin{figure}
\resizebox{13.7cm}{!}{\includegraphics*{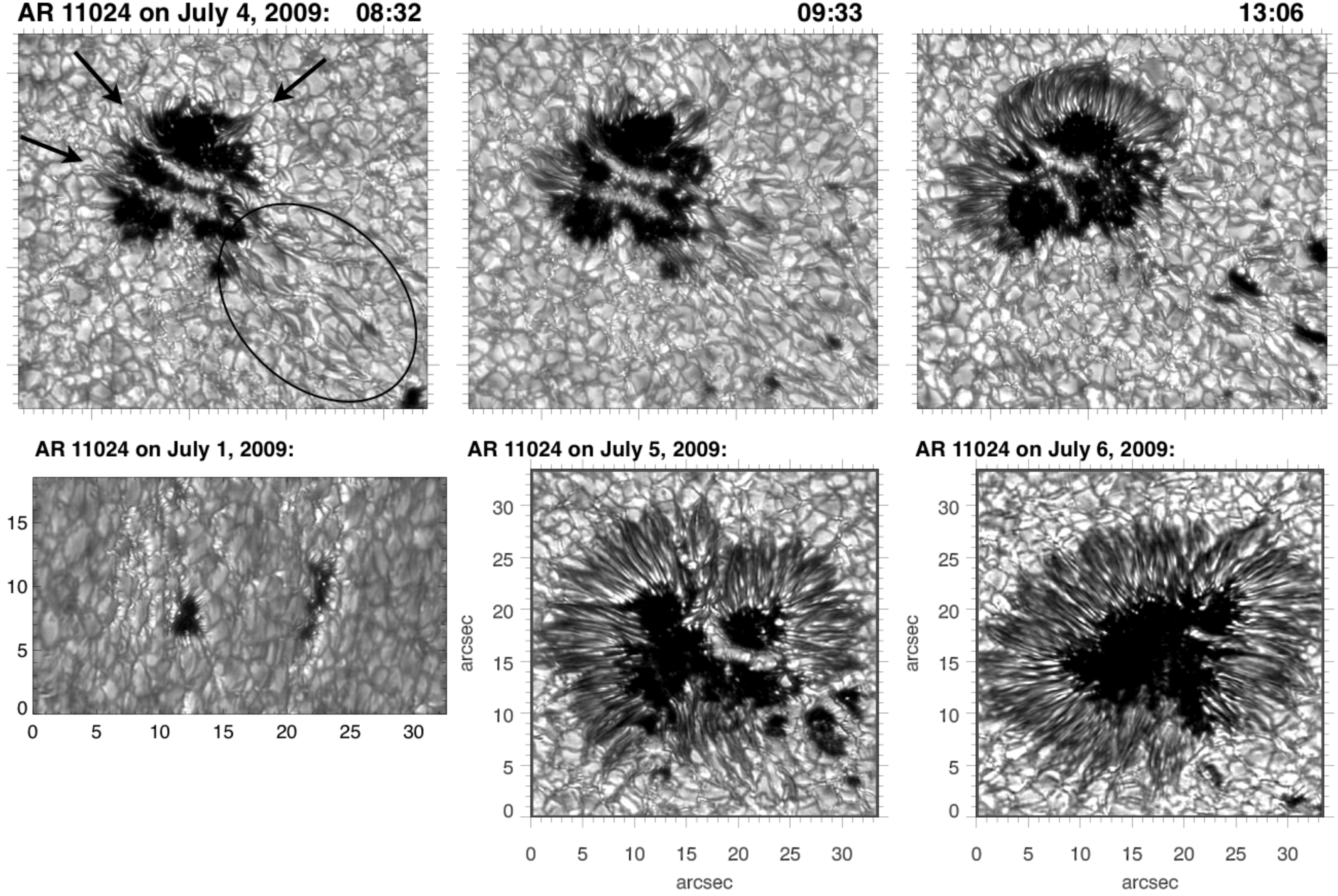}}
\caption{\label{fig:snapshots} Upper row: Three snapshot images in the G-band document the evolution of a sunspot penumbra in AR 11024 on July 4, 2009, at 08:32, 09:33, and 13:06 UT. 
Tick marks are in arcsec. Bottom row: Long term evolution of the active region is documented by an image of two pores of opposite polarity on July 1, and of the evolved spot on July 5 and July 6. Note the evolution of the light bridge which disappears on July 6.
}
\end{figure}

\begin{figure}
\resizebox{15.cm}{!}{\includegraphics*{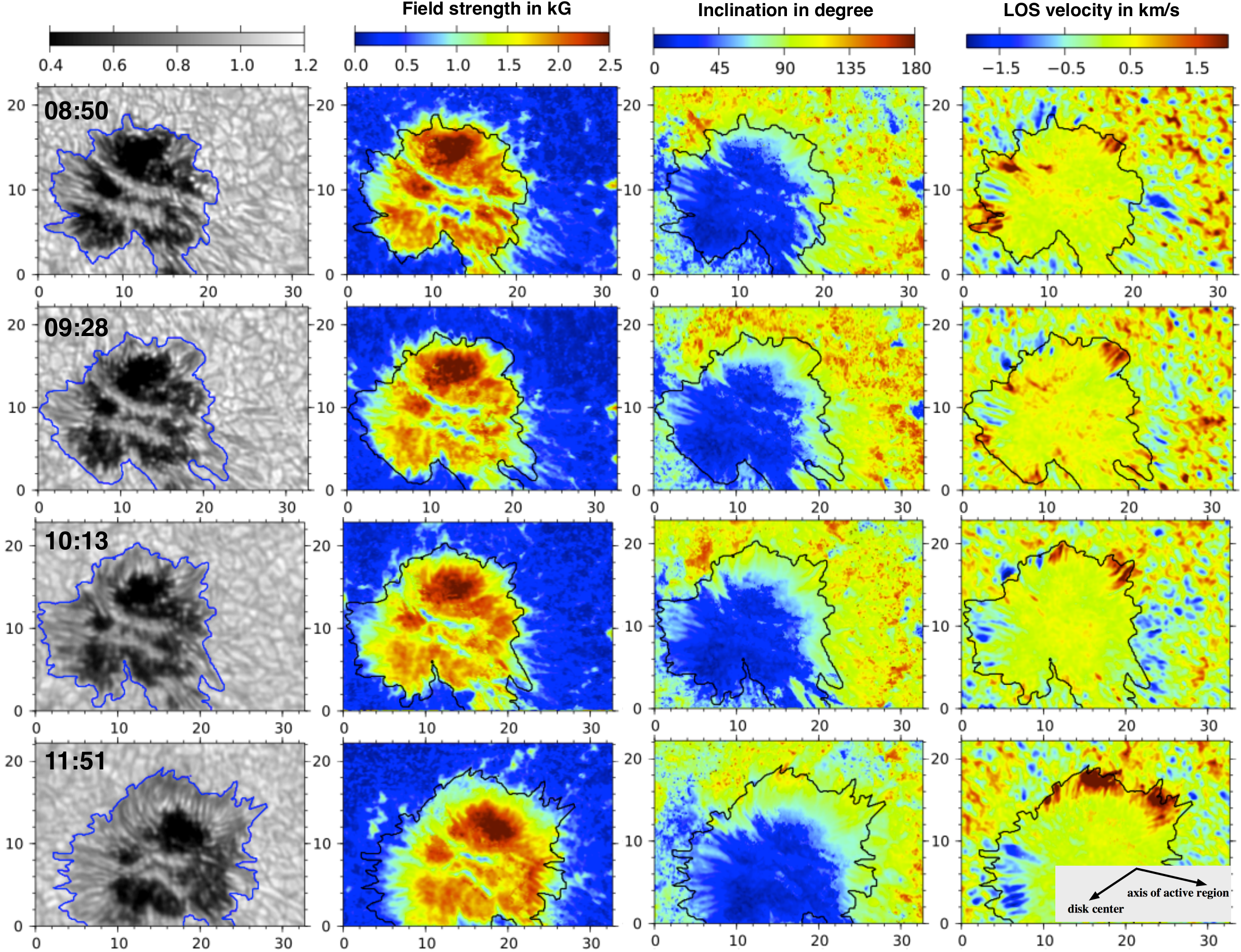}}
\caption{\label{fig:maps}Maps of AR 11024 for continuum, field strength, inclination, and LOS velocity at 08:50, 09:28, 10:13, 11:51.}
\end{figure}

\begin{figure}
\resizebox{14.5cm}{!}{\includegraphics*{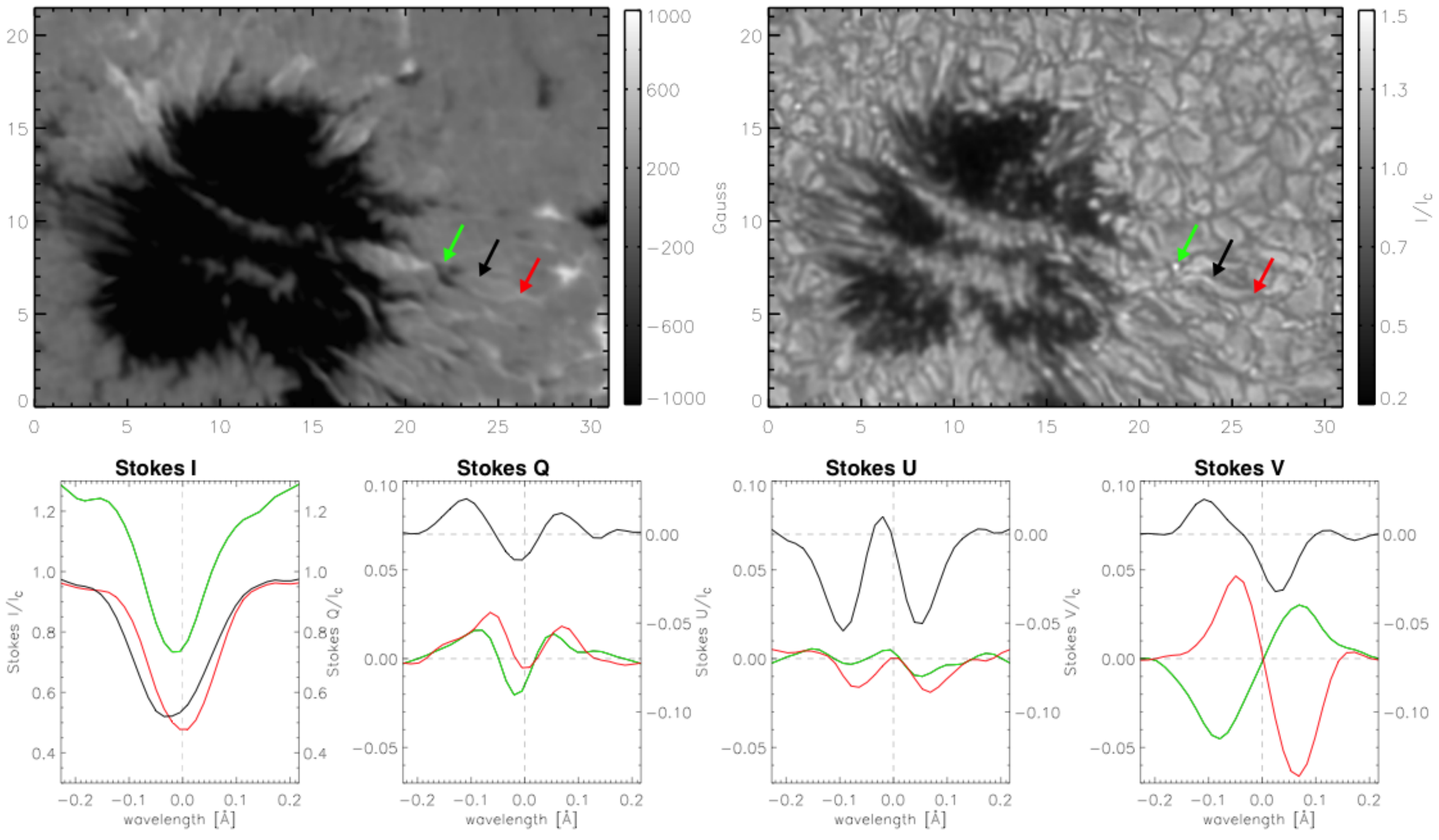}}
\caption{\label{fig:bipole}Upper row: Magnetogram and intensity image at 08:51 with GFPI. The colored arrows mark an elongated granule (emerging bipole). Lower row: Stokes profiles corresponding to the locations of the green, black, and red arrows, respectively. The Stokes V profiles of the green and red location are of opposite polarity.}
\end{figure}

\section{Penumbra Formation in AR 11024}
\label{sec:11024}

\subsection{Observations at the German VTT}

Between June 29 and July 10, 2009, we observed the disk passage of AR 11024 at the VTT \citep{schroeter+al1985}. Figure \ref{fig:snapshots} shows a selection of snapshots during its evolution, which are described in the course of this section.
As detailed in paper I, the campaign consisted of a multi-instrument setup and involved KAOS \citep[\underline{K}iepenheuer \underline{A}daptive \underline{O}ptics \underline{S}ystem, ][]{vdluhe+al2003}. Two imaging cameras were tuned to the G-band at 430 nm and to the Ca II K line core at 393 nm, respectively. These images were speckle reconstructed using KISIP \citep[\underline{K}iepenheuer \underline{I}nstitut \underline{S}peckle \underline{I}maging \underline{P}ackage, ][]{woeger+al2008}. This was enriched by two spectro-polarimeters of different type: the Fabry-P\'erot system GFPI \citep[G\"ottingen FPI, ][]{puschmann+al2006, bello+kneer2008} and the slit-based TIP \citep[\underline{T}enerife \underline{I}nfrared \underline{P}olarimeter, ][]{collados+al2007}. With the GFPI, we scanned Fe I 617.3 nm in 56 s to measure the maps of the Stokes parameters, and with TIP we observed Fe I 1089.6 nm, with an exposure time of 10 s per slit position.

When scanning the spot with TIP, the spot image also moves in the GFPI field-of-view. We made scans spanning not more than $2''$ with TIP to assure that the spot stays within the $30''$ by $20''$ field-of-view of GFPI. Yet, between UT 11:43 and 11:59 on July 4, we performed a scan with TIP covering the entire spot. 
With this scan we can demonstrate that the GFPI and the TIP not only measure similar maps of integrated Stokes profiles that look alike, but also produce physical maps that give consistent results for the two photospheric lines. 

For the GFPI and TIP data we performed one-component inversions with SIR \citep{ruiz+deltoro1992, bellot2003} to infer the physical parameters imprinted on the Stokes profiles by the Zeeman effect and other effects of radiative transport. To minimize the degrees of freedom, we assume that the magnetic and velocity fields are constant along the line-of-sight. Thereby, we correspondingly retrieve mean values, instead of attempting to resolve variations along the line-of-sight and within the resolution element. The data taken simultaneously with TIP and GPFI produce compatible inversion results. 

\begin{figure}
\resizebox{\textwidth}{!}{\includegraphics*{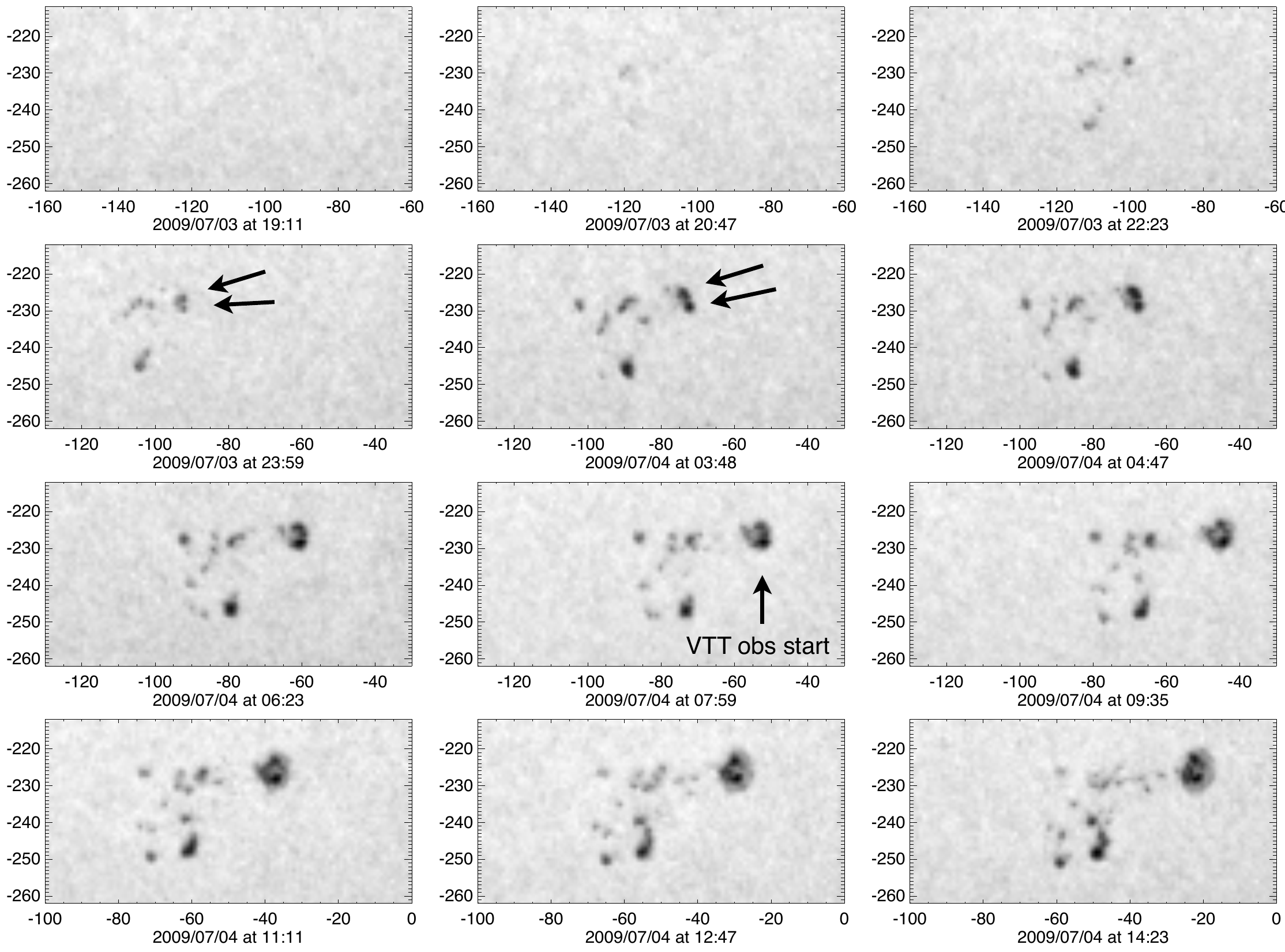}}
\caption{\label{fig:mdi:images}MDI continuum images of the active region starting at 19:11 on July 3 until 14:23 on July 4. The date and time are given below each image. The axis units are in arcsec and relative to disk center. X-axis varies to keep the active region in the field-of-view. The two adjacent arrows point at two forming pores which grow and merge to form the proto-spot. The single arrow at 07:59 mark the spot that we observed at the VTT, starting at 08:15.}
\end{figure}

\begin{figure}
\resizebox{\textwidth}{!}{\includegraphics*{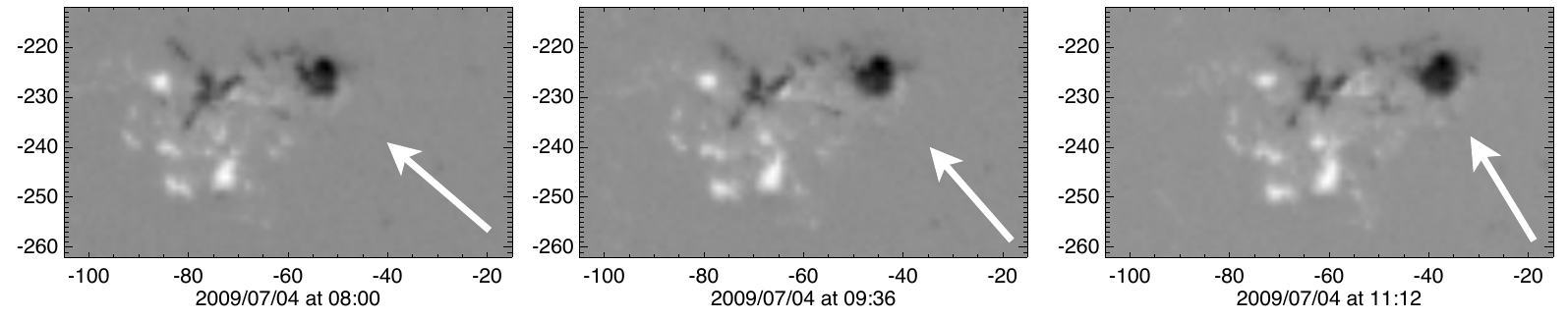}}
\caption{\label{fig:mdi:magnetograms}MDI magneto-grams at 08:00, 09:36, and 11:12 on July 4. The black-polarity leading spot is followed by some more diverged white polarity. Yet, the black polarity is not fully contained in the spot as there is also some black polarity flux in the emerging site between the main bulk of the polarities.}
\end{figure}

\subsection{Findings}

The upper row of Fig.~\ref{fig:snapshots} shows three speckle reconstructed snapshots at 08:32, 09:33, and 13:06 UT of our 4.5 hour time series on July 4, when the spot was at a heliocentric angle of 28 degrees and close to the meridian. In the lower right of the images (marked by the ellipse in the first frame) elongated granules reflect the emergence site of bipoles. Opposite to the emergence site, the penumbra forms. This formation starts from three locations marked by arrows in the upper left panel. Two of them are clearly connected to a light bridge. A close inspection with different intensity clippings revealed that the upper right formation area is also connected to a light bridge of smaller intensity. From these first filaments the penumbra forms in sections on a time scale of about 1 hour. The sections join and at the end of our time series they encircle more than half of the umbra (some 220 degrees). On the lower right side, where flux emergence continuously takes place, the spot does not form a stable penumbra, but only individual filaments which exist for some 10 minutes. 

In Fig.~\ref{fig:maps} we display 4 sets of inverted GFPI maps to note a couple of interesting features: (1) At 08:50, the field strength outside the continuum boundary of the spot is significant, while at 11:51 the boundary of the spot in continuum is very close to the boundary in magnetic field strength. (2) Penumbral filaments and penumbral areas are associated with large inclinations. The inclinations that are visualized in the third column of Fig.~\ref{fig:maps} are inclinations of the magnetic field vector relative to the line-of-sight.  (3) The map at 11:51 shows a 'regular' Evershed flow pattern. Most interestingly flow regions with opposite signs exist in early phases of the penumbra formation. (4) the magnetic field strength in the light bridges increases with time. According to VTT observations, only 3 days later, the light bridges also disappear in the intensity maps.

\subsubsection{Emerging Bipoles and Elongated Granules}

The emerging site is characterized by elongated granulation and intense proper motions of such granules towards and away the spot. These elongated granules play a crucial role for the penumbra formation: Such granules are associated with emerging bipoles (cf. paper II and Fig. \ref{fig:bipole}). The bipole axes, as well as the axes of their elongated granules, are oriented radially away from the center of the spot. All bipoles that we measure pop up in the photosphere such that the pole (footpoint) with the polarity of the spot is directed towards the spot. As the granule evolves, this pole migrates towards the spot, while the pole of the other polarity migrates away from the forming spot.

The example shown in Fig.~\ref{fig:bipole} shows an elongated granules which has strong field concentrations on either end. As seen in the lower right panel the recorded Stokes V profiles, these concentrations are of opposite polarity, and the concentrations closer to the spot has the polarity of the spot. From our time series we see that the black polarity is approaching the spot and finally merges with it (cf.~Paper I). The white polarity migrates away from the spot.

During the penumbra formation process, no pore merges with the spot. The increase of magnetic flux must exclusively be due to merging of small-scale field concentration which come from emerging bipoles. This bipoles emerge such that they are roughly aligned with the line between the two polarities, and most conspicuously, the polarity closer to the spot is always the same polarity as the spot. The two poles are located at the ends of an elongated granule. 

\subsubsection{Area, Magnetic Flux, and Field Strength} 

The area of the spot increases from 230 arcsec$^2$ to 360 arcsec$^2$ in 4.5 hours. This increase is exclusively taken up by the formation of the penumbra. The combined area of umbra and light bridges stays constant during that time (cf., paper I). From the magnetic field strength and its inclination as inferred from the inversion, we can compute the magnetic flux in the plane perpendicular to the line-of-sight, commonly referred to as the longitudinal magnetic flux. We find that this flux increases from $1.6\times 10^{21}$\,Mx to $2.4\times 10^{21}$\,Mx, i.e. the spot increases by $8\times10^{20}$\,Mx in 4 hours (08:40 until 12:38). At this rate a sunspot of $10^{22}$\,Mx would be formed in 2 days.
Since no other pore merges with the spot while the penumbra forms, we assume that all the additional flux is supplied by granular scale magnetic elements, which originate in bipoles. We find that a typical magnetic flux value for an individual magnetic element amounts to some $2$ -- $3\times 10^{18}$\,Mx. That means that about 1-2 emerging bipoles per minute (of which the spot-polarity footpoints subsequently merge with the spot) are needed to account for the increase of the magnetic flux of the spot. 
As the penumbra forms, we track the magnetic field strength of the spot. We find that the mean umbral and penumbral field strength stays constant at 2.2\,kG and at 1.5\,kG, respectively. Averaging 100 pixels with the largest field strength of the umbra, we also find a constant value in time amounting to 2.7\,kG (Rezaei, Bello Gonz\'alez, \& Schlichenmaier, in preparation).

\subsection{Global (Large-Scale) Evolution of AR 11024}

From our VTT campaign we gathered data from June 29 until July 10 of AR 11024. It rotated in onto the disk  on June 29. At this stage, it was visible as a facular region without pores. Its magnetic signatures are seen in MDI magneto-grams, but only barely in MDI continuum images. On July 1, the region formed two pores (lower left panel of Fig.~\ref{fig:snapshots}). These two pores -- of opposite polarity, as we know from TIP observations at the German VTT-- evolved significantly and disappeared by July 2.  In the morning of July 4, at 08:30 UT, a proto-spot with two pronounced light bridges appeared close to the zero-meridian of the Sun (upper row of Fig.~\ref{fig:snapshots}). In the following 4.5 hours a penumbra formed around the proto-spot. By 13:00 UT the seeing conditions did not allow for further observation. At that time the penumbra encircled some 220 degrees. No stable penumbra formed towards the site of magnetic flux emergence, i.e., in the direction of the opposite polarity of the active region.

Subsequently, the spot further increased in size. On July 5 the spot still exhibits a light bridge (lower mid panel of Fig.~\ref{fig:snapshots}). This light bridge disappeared by July 6 (lower right panel of Fig.~\ref{fig:snapshots}). Also then, the penumbra did not fully encircle the umbra, but had a gap of 30 degrees toward the opposite polarity of the active region. The spot further evolved, exhibiting a light bridge again two days later, and rotated off the solar disk after July 10. 

\subsubsection{Formation of the Proto-Spot}

Taking advantage of the MDI data base \citep{scherrer+al1995}, we can track the formation of the proto-spot with MDI full disk continuum images. A series of images is shown in Fig.\,\ref{fig:mdi:images} spanning the time between 19:11 on July 3 until 14:23 on July 4. While no pores are visible on the first image, some dark patches are seen at 20:47 on July 3. At 22:23, there is a pronounced pore leading the active region (upper right), being located at $(-100'',-225'')$. In the first image of the second row (23:59) a second leading pore has appeared. Now the pores are at $(-92'',-230'')$, and the magneto-grams in Fig.\,\ref{fig:mdi:magnetograms} indicate that they are of the same polarity. These pores are already very close to each other since their first appearance. Rather than migrating towards each other, it appears that they increase in size, and the granulation between them transforms into a light bridge. At about 07:59 (middle image of third row) they reach a state that can be described as one object, i.e., a proto-spot has formed. Our observations at the VTT started at 08:30. At this stage the proto-spot started to develop penumbral segments.

The formation of the penumbra is also visible in the MDI images (bottom two rows of Fig.\,\ref{fig:mdi:images}). It is seen that the penumbra forms on the side facing away from the active region. \citet{zwaan1992} -- refering to \citet{bumba+suda1984} and \citet{mcintosh1981} -- ascribes the formation of a sunspot to the coalescence of the existing pores. The case that we witness here is consistent with this picture, although the pores do not merge by approaching each other, but form very close to each other and further increase in size. 

\begin{figure}
\resizebox{15cm}{!}{\includegraphics*{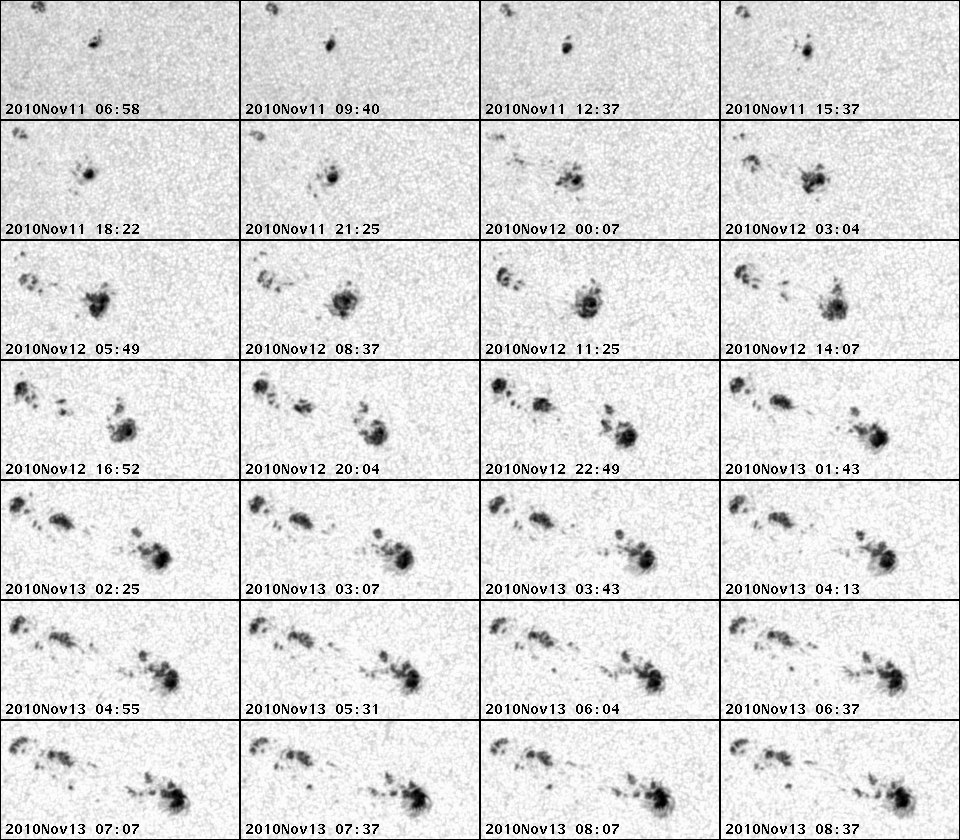}}\\[-1pt]
\resizebox{15cm}{!}{\includegraphics*{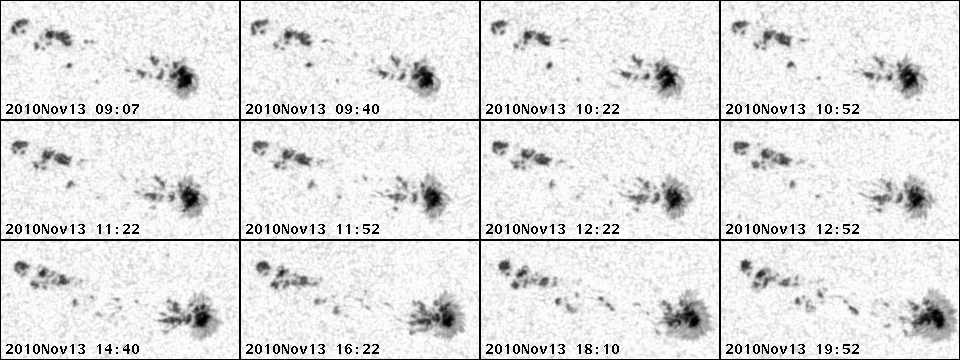}}
\caption{\label{fig:11124} Evolution of AR 11124 as observed with HMI. During the phase of the formation of the penumbra (Nov 13, 01:43 until 12:52) the time steps are reduced. The field-of-views are 120x60 arcsec$^2$. On Nov 12, 2010, 20:04 (4th row, 2nd column), the heliocentric angle of the lower right spot (the leading polarity) amounts to about 16 degree.
%
}
\end{figure}

\begin{figure}
\resizebox{13cm}{!}{\includegraphics{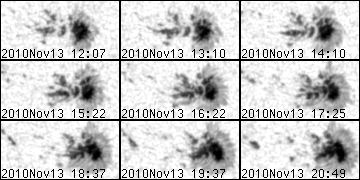}}
\caption{
\label{fig:pore} Evolution of the spot in AR 11124 on Nov 13, 2010, as a pore merges. As a consequence of the merging pore, a light bridge develops within the spot. This initial light bridge disappears in the course of the further evolution.
}
\end{figure}

\section{Observations with HMI Onboard SDO}
\label{sec:11124}

Since the beginning of 2010, HMI monitors the full Sun at a spatial resolution of about 1 arcsec with 4k by 4k pixel cameras. This data is ideally suited to study the formation of active regions and the formation of sunspots. Although HMI records spectro-polarimetric data in Fe I 617.3 nm, here we only use continuum images from HMI downloaded from \verb+ http://sdo.gsfc.nasa.gov/+\footnote{From that web page, 4k by 4k HMI continuum images appear to be of better spatial resolution than the AIA 4500 images.}. 
We use these images to document the formation of AR 11124, and to compare that particular event with the formation of AR 11024 from the preceding section. There are many other examples of sunspot formation recorded with HMI, but it is beyond the scope of this paper to present a systematic analysis of sunspot formation.

Figure \ref{fig:11124} displays HMI continuum images with a field-of-view of 120 by 60 arcsec$^2$ monitoring the formation of AR 11124 and the development of the penumbra in the leading spot. The active region is in the northern hemisphere and on Nov 12, 20:00, the leading spot has a heliocentric angle of some 16 degrees.

The proto-spot forming phase of AR 11124 is similar to AR 11024: pores form and disappear prior to the main emergence phase. The latter starts on Nov 11: In the morning two pores of opposite polarity have formed (first row in Fig. \ref{fig:11124}). In the further course of that day several small pores form and create a halo around the leading spot. Elongated structure starts to become visible between the two polarities, see, e.g., Nov 12, 00:07. During  Nov 12 all the pores of the leading polarity clump together and form the proto-spot which starts to develop a penumbra at 01:43 on Nov 13. During this formation phase, movies give the impression that the spot rotates. This impression is given by twisted -- i.e., non-radial -- penumbral filaments, which rotate and eventually become radial. This happens at the end of the penumbra formation phase at about 14:40 (Nov 13). The penumbra forms in segments, except on the side facing the emergence site. There, the ongoing activity apparently inhibits the penumbra formation. At 14:40 the penumbra formation process can be considered as being completed, hence the formation took about 12 hours. 

The penumbra continues to evolve and to grow in size as a result of subsequently merging pores. In this AR 11124, pores from large distance migrate toward and merge with the spot. In AR 11024, this is not observed. There the two pores that form the proto-spot are very close to each other from the very beginning, and later, the magnetic flux is increased by small scale elements.

\paragraph{Merging Pores and Light Bridges} Figure \ref{fig:pore} monitors a rather typical example of a pore migrating towards and merging with the spot. The figure shows nine snapshots between 12:07 and 20:49 on Nov 13.
As the pore approaches the spot, the bright lane between pore and spot becomes narrower. At 16:22, pore and spot have merged and the 'granulation' in between becomes visible as a light bridge. This light bridge disappears in the further course of the spot evolution. It is almost invisible 3.5 hours later at 19:52. The pores that merge with the spot in the afternoon of Nov 13 seem to have increased the magnetic flux considerably with the effect that both, the umbra and the penumbra become larger. At this time the penumbra is large, but -- as it is typical for forming sunspots -- the penumbra still does not encircle the full penumbra. Towards the emergence site no penumbra forms. This is a common feature of AR 11024 and AR 11124.
There is another very interesting common feature: In both cases the process of a merged pore produces light bridges within the spot. And in both cases the light bridges disappear after the formation phase. 

\paragraph{Emerging Bipoles} Elongated granules are visible in a number of HMI snapshots. They are known to trace emerging bipoles. It seems straight forward to surmise that the magnetic elements from the emerging bipoles is gathered to form numerous pores observed during the formation of AR 11124. In turn, the spot forms by merging pores and by merging magnetic elements. Therefore both, AR 11024 and AR 11124, seem to indicate that emerging bipoles are the building blocks of larger structures like pores and sunspots.

\section{Conclusions} \label{sec:summary}

At the VTT on July 4, 2009, we observed a region of emerging magnetic flux in AR 11024, in which a proto-spot without penumbra forms a penumbra within some 4.5 hours. This process is documented by multi-wavelength spectro-polarimetric observations at a spatial resolution as good as 0.3 arcsec. MDI continuum maps and magnetograms are used to follow the formation of this particular proto-spot, and the subsequent evolution of the entire active region.

We find that the proto-spot of AR 11024 forms by merging of two pores that increase in size. During the formation phase of the penumbra no further pore joins the spot. Instead magnetic elements join the spot. They come from bipoles that emerge in between the two AR polarities. Thereby the magnetic flux increases, until a critical value seems to be reached at which the spot changes its field topology: segments of penumbrae form, preferably at the junctions between light bridges and surroundings. Yet, the penumbra avoids the side directed towards the site of emergence. This indicates that the magnetic field needs space and some degree of quietness to change its configuration into the penumbral type. 

We cannot give a definite answer on the question why the penumbra forms the way it does, except for the canonical answer that the field inclination at the outer boundary of a pore increases with increasing magnetic flux, and that at some critical inclination angle the mode of magneto-convection changes from umbral to penumbral, i.e., from vertical mean fields into inclined mean magnetic fields. Concerning the flow field, we observe peculiar flows prior to the formation of penumbral segments. These flows have the opposite direction as the regular Evershed flow. Yet, we do not understand why nor how these flows trigger the onset of penumbral formation. 

As a summary, we conclude that sunspots form out of merging pores and magnetic elements.  Light bridges in forming sunspots are remnants of merging pores, and typically disappear in the further course of the evolution.
The building blocks of active regions are emerging bipoles. They organize themselves to form larger structures: pores and sunspots. The ordered process of reorganization is a strong indication for a common root of each polarity. The penumbral mode of magneto-convection starts at the outer end of light bridges, but only if the magnetic field can spread out without getting disturbed by ambient fields, and in particular without getting pushed by incoming flux. As soon as penumbral filaments are developed the regular Evershed flow is present and the magnetic field is largely inclined.

\paragraph{Acknowledgements}
The German VTT is operated by the Kiepenheuer-Institut f\"ur Sonnenphysik at the Observatorio del Teide in Tenerife. We acknowledge the support by the VTT and KAOS group. NBG acknowleges the Pakt f\"ur Forschung. SOHO is a project of international cooperation between ESA and NASA. For the HMI data we thank NASA/SDO and the AIA, EVE, and HMI science teams.


\end{document}